\documentclass[RNAAS]{aastex631}

\newcommand{\halpha}{H-$\alpha$}

\shorttitle{V 1298 Tau c H-Alpha Variability}
\shortauthors{Schlawin et al.}
\graphicspath{{./}{figures/}}
\begin{document}

\title{H-Alpha Variability of V1298 Tau c}

\author[0000-0001-8291-6490]{Everett Schlawin}
\affiliation{Steward Observatory}% \\
%933 North Cherry Avenue \\
%Tucson, AZ 85721, USA}

\author[0000-0002-0551-046X]{Ilya Ilyin}
\affiliation{Leibniz-Institute for Astrophysics Potsdam (AIP)}
%\\
%An der Sternwarte 16, 14482 Potsdam, Germany}

\author[0000-0002-9464-8101]{Adina D. Feinstein}
\affiliation{Department of Astronomy and Astrophysics, University of Chicago}
\affiliation{NSF Graduate Research Fellow}
%
%5640 South Ellis Avenue, Chicago, IL 60637, USA}

\author[0000-0003-4733-6532]{Jacob Bean}
\affiliation{Department of Astronomy and Astrophysics, University of
Chicago}%, 5640 South Ellis Avenue, Chicago, IL 60637, USA}

\author[0000-0001-9446-6853]{Chenliang Huang}
\affiliation{Lunar and Planetary Laboratory}%, University of Arizona, Tucson, Arizona 85721, USA}

\author[0000-0002-8518-9601]{Peter Gao}
\affiliation{Earth and Planets Laboratory, Carnegie Institution for Science}%\\
%5241 Broad Branch Road, NW, Washington, DC 20015, USA}

\author{Klaus Strassmeier}
\affiliation{Leibniz-Institute for Astrophysics Potsdam (AIP)}%\\
%An der Sternwarte 16, 14482 Potsdam, Germany}

\author[0000-0003-1231-2194]{Katja Poppenhaeger}
\affiliation{Leibniz-Institute for Astrophysics Potsdam (AIP)}%\\
%An der Sternwarte 16, 14482 Potsdam, Germany}
\affiliation{University of Potsdam, Institute of Physics and Astronomy}%\\
%Karl-Liebknecht-Str. 24/25, 14476 Potsdam, Germany}



\begin{abstract}

The 23 Myr system V1298 Tau hosts four transiting planets and is a valuable laboratory for exploring the early stages of planet evolution soon after formation of the star.
We observe the innermost planet, V1298 Tau c, during transit using LBT PEPSI to obtain high spectral resolution characterization of escaping material near the H-$\alpha$ line.
We find no strong evidence for atmospheric material escaping at the orbital velocity of the planet.
Instead, we find a deep stellar feature that is variable on the few percent level, similar to a previous observation of the planet and can be explained by stellar activity.
We attempted to monitor the broadband optical transit with LBT MODS but do not achieve the precision needed to characterize the atmosphere or improve the ephemeris.

\end{abstract}

%\keywords{Exoplanet astronomy (486) --- Exoplanet atmospheres (487) --- Stellar activity (1580) --- Planet formation (1241) --- Exoplanet evolution (491)}


\section{Introduction} \label{sec:intro}

Statistical surveys of planet occurrence rates indicate that there is a radius gap between Earth-sized and sub-Neptune-sized planets \citep{fulton2017radiusGap} that could arise from core-powered mass loss \citep{gupta2019MNRAS.487...24G} and/or photoevaporation \citep{owen2017evaporationValley}.
%Studying newly formed and young systems can better increase our understanding of how the atmospheric mass may evolve and carve out this gap.
V1298 Tau is a young (23 Myr) bright system with 4 transiting planets \citep{david2019V1298TauPb,david2019V1298TauPcde} that can reveal the evolutionary processes shaping the radius gap.
%The young age of the system provides a window into the early evolutionary stages of a planet system and the transiting geometry reveals the physical radii of the planets.
%The transiting nature of the planets also permits characterization of their sizes, densities, atmospheric composition, and whether they are currently undergoing mass loss.
We observe the V1298 Tau c planet to search for escaping Hydrogen and to constrain the atmospheric hazes or clouds with optical broadband spectroscopy.

\section{Observations}\label{sec:obs}
We observed V1298 Tau c with the Large Binocular Telescope (LBT) on UT 2020-11-14 during primary transit under LBT program AZ-2020B-016.
We used an ephemeris derived from Spitzer Data (John Livingston, private communication).
% with a time of transit center $T_0$=2458846.097 and a Period of 8.249143$^{+0.000026}_{-0.000036}$ days.
The telescope was configured in heterogenous binocular mode where light from the DX 8.4 m primary fed the PEPSI instrument while light from the SX 8.4 primary fed the MODS-1 instrument.

The PEPSI instrument was configured to use the CD5 disperser on the red side to capture the 656 nm \halpha\ line and the CD1 disperser on the blue side to capture the Ca II H and K lines at 393.3 and 396.8 nm, with a resolution of R$\approx$120,000.
%The 200~\micron\ fiber was used to give a resolution of R$\approx$120,000.
%The wavelengths covered were therefore 384 to 427 nm and 628 nm to 742 nm, which did not permit characterization of the H-$\beta$, Mg I triplet, He I, Ca II infrared triplet, Na I nor K I that are sometimes explored in exoplanet atmospheres \citep[e.g.][]{hoeijmakers2019kelt9Survey,feinstein2021v1298taucCaHalpha}.
%The PEPSI instrument collected exposures at a cadence of approximately 10.8 minutes.

Standard pipeline processing of the PEPSI spectra was used with the numerical toolkit and graphical interface described in \citet{strassmeier2018sunPhotosphere}.
%The PEPSI data processing steps included bias subtraction and variance estimation of the source frames, master flatfield correction for the CCD spatial noise, echelle order definition from the flats, scattered light subtraction, wavelength solution for the ThAr frames, optimal extraction and cosmic spikes elimination of the target frames, wavelength calibration, normalization to the master flat field spectrum to remove CCD fringes and blaze function, a global 2D fit to the continuum of the normalized image, and rectification of all spectral orders in the frames to a 1D spectrum.
%Remaining
Cosmic ray outliers above 5\% from the continuum were removed from the 1D spectrum and were replaced by linear interpolation of neighboring points.
We find that the Ca II H\&K lines are seen in emission, but have too low signal-to-noise (SNR $\approx$ 7) to adequately study variability near these lines.

\section{Results}\label{sec:results}
\begin{figure*}[ht!]
\gridline{
	\fig{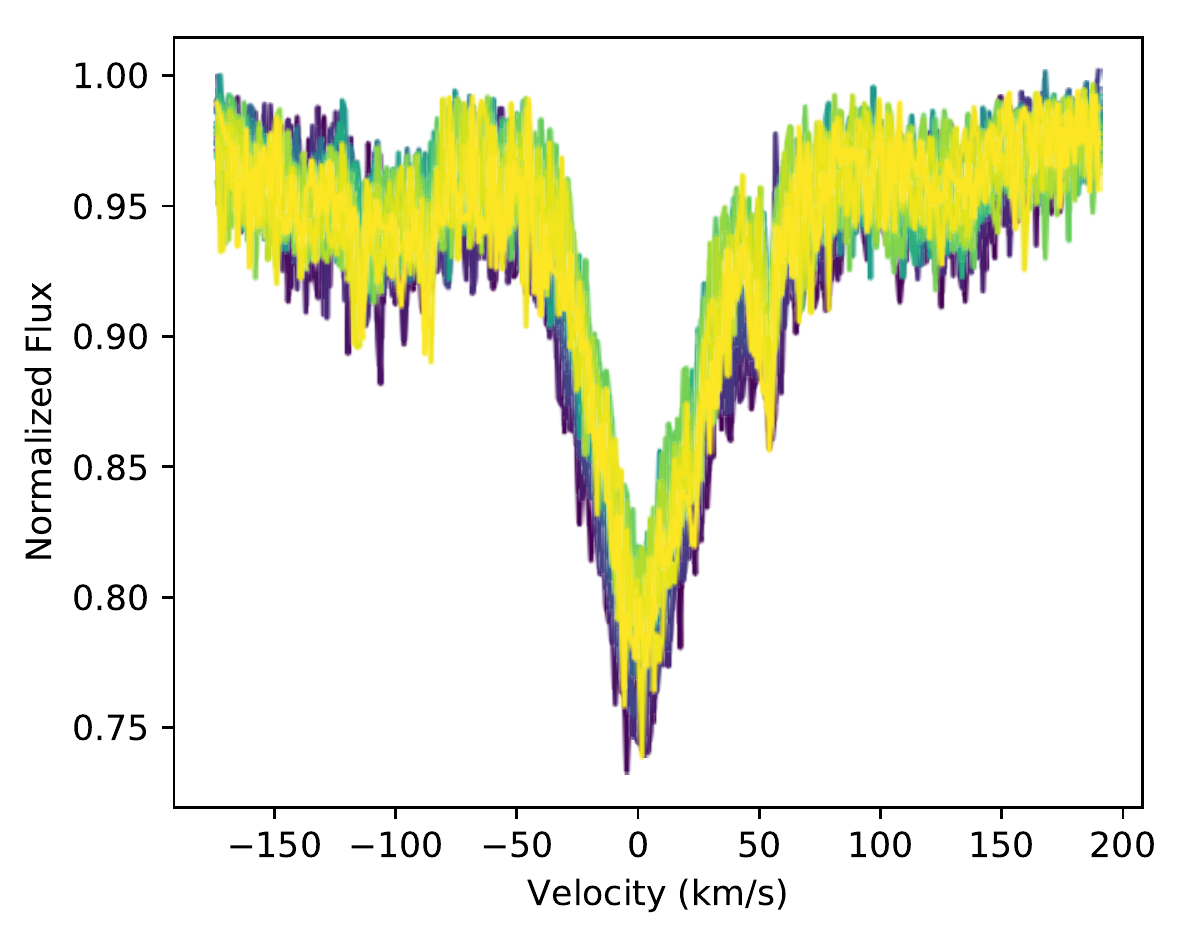}{0.33\textwidth}{\halpha\ Line Profile}
	\fig{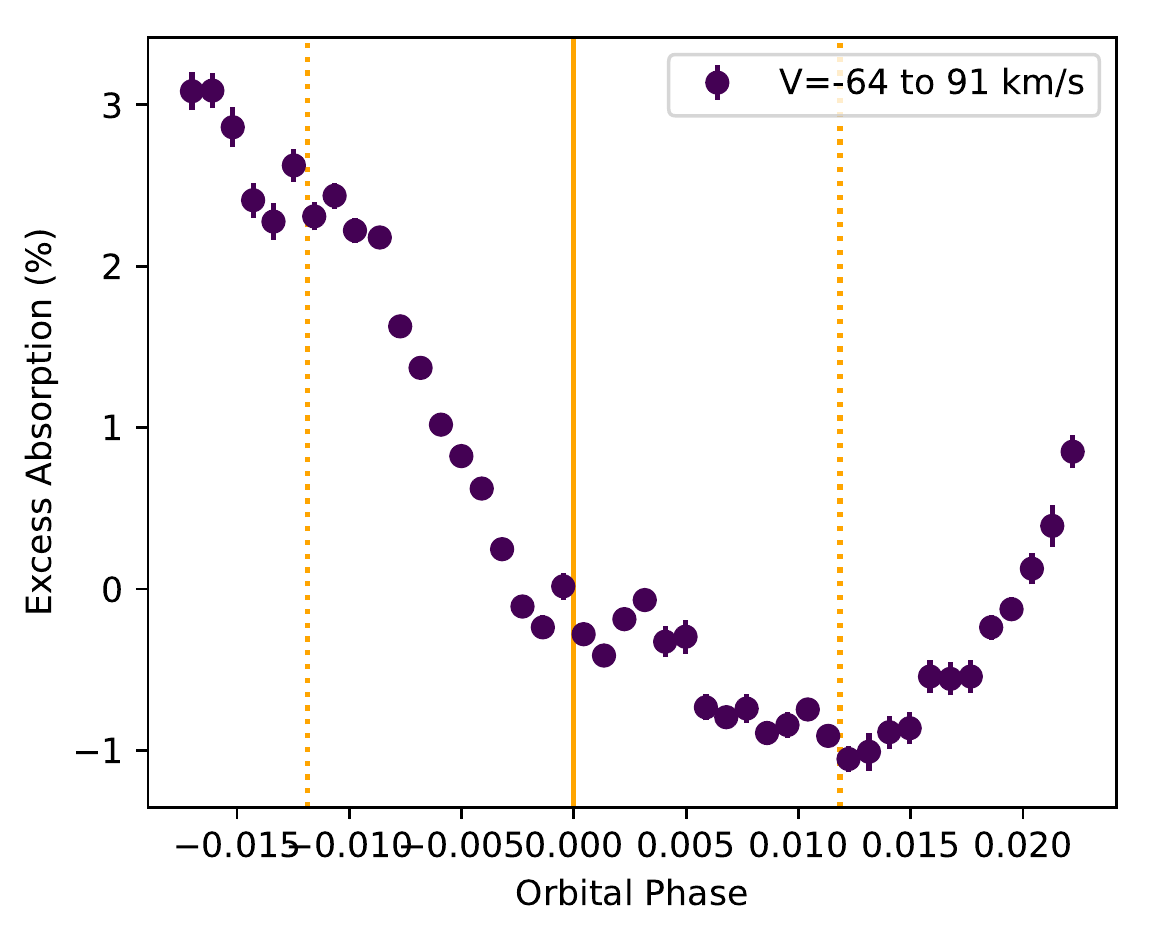}{0.33\textwidth}{Absorption Time Series}
	\fig{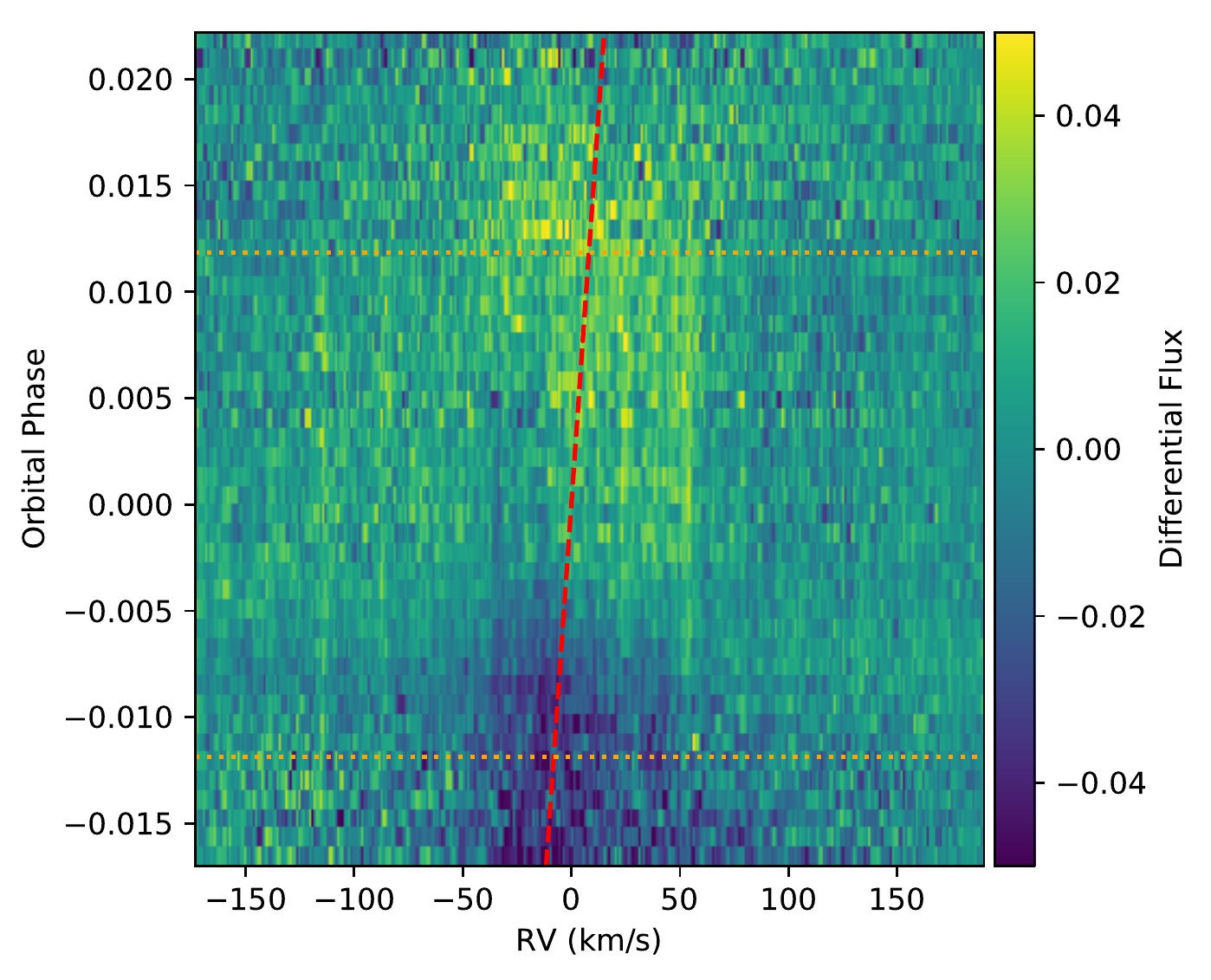}{0.33\textwidth}{Dynamic Spectrum}
	}
\caption{The \halpha\ absorption line (left) is variable through the transit (middle).
The fractional deviation from the mean of all out-of-transit spectra (right) shows that the absorption does not occur at the planet rest velocity (red dashed line) and is instead likely related to stellar activity.
The absorption is also not confined to the transit which is between first and fourth contact (orange dotted lines).
\label{fig:general}}
\end{figure*}

We find an \halpha\ profile that is similar to \citet{feinstein2021v1298taucCaHalpha}.
There is excess absorption at 55 km/s due to telluric contamination.
The \halpha\  line is variable, which we characterize with a velocity-integrated summation using the same velocity width as \citet{feinstein2021v1298taucCaHalpha}: -64 to 91 km/s.
%We compare this band-integrated depth to the median out-of-transit value and find that there is a decrease through the transit followed by an increase after egress.
We show the time series of the absorption in Figure \ref{fig:general}, where the absorption decreases toward-mid-transit and increases after egress.
Here, we plot the excess absorption, defined as
\begin{equation}
F_{excess} = \frac{ F_{mean} - F(t)}{F_{mean}},
\end{equation}
where $F(t)$ is the band-integrated flux from -64 to 91 km/s and $F_{mean}$ is the mean value of $F(t)$ for the out-of-transit measurements.
We compare the variability to the projected orbital velocity of the planet as seen in the differential dynamic spectrum in Figure \ref{fig:general}.
For the orbital parameters, we use the inclination and semi-major axis from \citet{david2019V1298TauPcde} and the Spitzer ephemeris.
The \halpha\ variability does not follow the projected orbital velocity of the planet.

We note that the \halpha\ variability direction and magnitude is very similar to \citet{feinstein2021v1298taucCaHalpha}.
This could be due to similar chromospheric star spot and faculae geometry of the star at the two epochs.
%The PEPSI observations we observed here occurred on UT 2020-11-14 compared to UT 2020-01-22 for the GRACES spectroscopy in \citet{feinstein2021v1298taucCaHalpha}.
%It is also possible that the activity is related to the planet V1298 Tau c due to star-planet interactions \citep[e.g][]{shkolnik2003}.
%However, the variability of the \halpha\ line is consistent with other young stars in general \citep{feinstein2021v1298taucCaHalpha} so it is likely completely stellar.
Given that the projected rotation velocity of the star is 23~km/s \citep{david2019V1298TauPb}, some of the \halpha\ variability could be related to faculae appearing in the line of sight as the star rotates.
At an orbital phase near 0.013, there is excess flux at about -20~km/s that could be expected if faculae appear on the approaching limb of the star as it rotates.

Finally, we report that the MODS spectra did not give sufficient precision for characterization of the atmosphere.
The MODS spectra were obtained in prism mode and were analyzed by both box extraction and multi-object spectroscopy but both methods led to large variability in the time series.

Continued monitoring of the system will provide insight about the atmospheres of the four known planets as well as the stellar variability.
Narrowband photometry of the He line suggests that V1298 Tau d has an absorption signature while V1298 Tau c does not \citep{vissapragada2021v1298tauHe}.

%% The "ht!" tells LaTeX to put the figure "here" first, at the "top" next
%% and to override the normal way of calculating a float position

%\begin{figure}[ht!]
%\centering
%\includegraphics[width=.99\columnwidth]{v1298comb_fig.pdf}
%%\plotone{}
%\caption{{\it Left:} H-$\alpha$ line profile. {\it Middle:} Absorption Time Series.
%{\it Right:} Dynamic Spectrum  \label{fig:general}}
%\end{figure}

\begin{acknowledgments}
Funding for E Schlawin is provided by NASA Goddard Spaceflight Center.
We respectfully acknowledge the University of Arizona is on the land and territories of Indigenous peoples. Today, Tucson is home to the O'odham and the Yaqui.
%Committed to diversity and inclusion, the University strives to build sustainable relationships with sovereign Native Nations and Indigenous communities through education offerings, partnerships, and community service.
ADF acknowledges the support from the National Science Foundation Graduate Research Fellowship Program under Grant No. (DGE-1746045). 
We appreciate the valuable observation planning and execution of the complicated heterogeneous binocular mode by Jennifer Power and Olga Kuhn.
\end{acknowledgments}


\vspace{5mm}
\facilities{LBT(PEPSI), LBT(MODS)}


\software{astropy \citep{astropy2013}, 
          \texttt{photutils v0.3} \citep{bradley2016photutilsv0p3},
          \texttt{ccdproc} \citep{craig2015ccdproc}
          \texttt{matplotlib} \citep{Hunter2007matplotlib},
          \texttt{numpy} \citep{vanderWalt2011numpy},
          \texttt{scipy} \citep{virtanen2020scipy},
          }



%\appendix


\bibliography{this_biblio}{}
\bibliographystyle{aasjournal}



\end{document}